\documentclass[aps,pra,reprint,superscriptaddress,nofootinbib,longbibliography]{revtex4-2}

\usepackage{amsmath}
\usepackage{amssymb}
\usepackage{bm}
\usepackage{graphicx}
\usepackage{hyperref}

\newcommand{\vp}{\vec{p}}
\newcommand{\vx}{\vec{x}}

\begin{document}

\title{Entangled photons from classical field theory}

\author{C.~Wetterich}
\email{c.wetterich@thphys.uni-heidelberg.de}
\affiliation{Institut f\"ur Theoretische Physik, Universit\"at Heidelberg,
Philosophenweg 16, D-69120 Heidelberg, Germany}

\begin{abstract}
The violation of Bell's inequalities for a pair of entangled photons has been
interpreted as a no go theorem for the embedding of quantum mechanics into a
classical probabilistic setting. We demonstrate how these entangled photons
are described by a probabilistic classical field theory. The non-commuting
operators for polarization observables at different angles are related to
the presence of statistical observables in a classical probabilistic
setting. The experimentally observed correlations for entangled photons are
compatible with a classical probabilistic model. They point to the importance
of statistical observables.
\end{abstract}

\maketitle

\section{Introduction}\label{sec:intro}

Classical correlation functions have to obey Bell's
inequalities~\cite{BEL1,CHSH}. Experiment has established the violation
of these inequalities by polarization measurements for entangled
photons~\cite{FRCL,ASP1,ASP2,ASP3}. This is generally interpreted as a no-go
theorem for the embedding of quantum mechanics into classical statistics. This
paper provides a counter-example for such a no-go theorem. Based on classical
field equations for free electromagnetic fields with probabilistic initial
conditions we establish the presence of states with a finite number of
photons. They constitute subsystems of the probability distribution which are
closed under the time evolution. Photons with momentum $\vp=(0,0,p_3)$ have
two possible polarization directions perpendicular to $\vp$. The photon
numbers $n_1(\vp)$ and $n_2(\vp)$ for the two polarizations are conserved
separately. Our basic assumption on ideal polarization measurements states
that a polarization detector with filter in the one-direction measures for a
one-photon state the expectation value $\langle n_1(\vp)\rangle$. This holds
in the sense that the probability to detect the photon is given by
$\langle n_1(\vp)\rangle$, as computed from the classical probabilistic
information. Extending this to entangled two-photon states with polarization
filters at arbitrary angles $\alpha$ and $\beta$ we compute correlation
functions for the two photons. They agree with the quantum correlations and
violate the CHSH-inequality~\cite{CHSH}. We conclude that a classical
probabilistic setting can produce correlations which violate Bell's
inequalities. The experimental finding of this violation can therefore not be
taken as an argument that an embedding of quantum mechanics into a classical
probabilistic system is not possible.

The central reason for the violation of Bell's inequalities is incomplete
statistics~\cite{CWIS1,CWIS2,CWPW}. Bell's inequalities can be derived for
complete classical statistics, which assumes that for arbitrary pairs of
observables $A$ and $B$ one can define simultaneous positive probabilities
$p(\lambda_A,\lambda_B)$ for every pair of possible measurement values
$\lambda_A$ for $A$ and $\lambda_B$ for $B$. Classical correlation functions
exist only if this completeness condition is met. Indeed, the CHSH
inequalities hold for all settings if and only if a joint probability
distribution for the four observables involved exists~\cite{FINE}. For the polarization
observables of the two-photon system we find that the completeness condition
is not obeyed simultaneously for all pairs of angles of the polarization
filter. Correlations for arbitrary polarization measurements for the
two-photon subsystem can be derived naturally from our basic assumption for a
single-photon polarization detector. They are not classical correlation
functions, however.

In our case incomplete statistics is related to the presence of statistical
observables~\cite{CWPW,CWCPT,CWQF,CWQP}. Statistical observables measure
properties of the probabilistic information, conceptually somewhat similar to
temperature, pressure or local magnetization in macroscopic classical
equilibrium systems. For example, the quantum angular momentum is a
statistical observable which measures properties of the probabilistic
information under rotations. In our setting the conserved photon numbers are
key statistical observables. They reflect the presence of subsystems with a
finite conserved number of photons. Statistical observables are not functions
of the classical fields. They are typically associated to operators which also
involve derivatives with respect to the fields. The operators for statistical
observables typically do not commute. In this case simultaneous probabilities
$p(\lambda_A,\lambda_B)$ are not possible and the statistics is incomplete.

Statistical observables often arise from projections to
subsystems~\cite{CWPW}. The polarization observables for one- or two-photon
systems are such subsystem observables. In our approach the individual
photons should be seen as excitations of a vacuum state. They are not
isolated objects. The classical probability distribution for the ``quantum
vacuum'' is a Gaussian in the field variables with non-zero width. This
contrasts with the classical vacuum which is a sharp distribution for
vanishing field values. The properties of the quantum vacuum enter the
dynamics and correlations of the photon excitations.

An important concept for the understanding of statistical observables is the
real classical wave function $q(\tilde\sigma)$~\cite{CWQP1,CWIT,CWQFC,CWQF}.
It is the square root of the probability distribution,
$w(\tilde\sigma)=q^2(\tilde\sigma)$. Both $w$ and $q$ are functions of the
classical field variables $\tilde\sigma$. Expectation values of linear
observables are formed as $\langle A\rangle=q^T\hat Aq$. For ``classical
observables'' $\hat A$ is a function of $\tilde\sigma$ and the rule for the
expectation value is the standard rule for classical statistics. The advantage
of the wave function is the possibility to compute also expectation values for
statistical observables for which $\hat A$ involves field derivatives. The
classical wave function differs from the complex
Koopman--von~Neumann--Sch\"onberg (KNS) wave
function~\cite{KOOP,VNEU,SCHB}. For the KNS wave functions the phases are
redundant and do not influence the dynamics or expectation values. We will
map $q$ to a complex wave function $\tilde\psi$, and the phases of $\tilde\psi$
will contain crucial information.

The many interesting developments of the
operator formalism based on the KNS-approach~\cite{KAN,MAMA,MAU,GOMA,NIC,NAKL,VOLO,BON,AME,KPM,ACKA,CHRU,KLE,MEZI,DW,GOZ1,GRT}
have been considered as a description of classical physics. Our approach with
the real classical wave function and operators for statistical observables
shows the presence of a true quantum formalism within a classical
probabilistic setting.

We begin in sect.~\ref{sec:field} with the classical field equations for the
electromagnetic gauge field. In sect.~\ref{sec:prob} we introduce the
classical probability distribution and wave function for field
configurations. We find a free quantum field theory for electromagnetism as a
subsystem. In sect.~\ref{sec:photons} we discuss the vacuum and photon
excitations. Sect.~\ref{sec:polmeas} formulates our basic assumption for a
polarization measurement of a single photon. In sect.~\ref{sec:Bell} we
discuss the correlations for two-photon systems and establish the violation
of the CHSH-inequality. Sect.~\ref{sec:statobs} provides for a more general
discussion of statistical observables and our conclusions are found in
sect.~\ref{sec:concl}.

\section{Classical field equations}\label{sec:field}

Our starting point is the gauge- and Lorentz-invariant classical field
equation for the four-component gauge field $A_\mu$ in vacuum,
\begin{equation}\label{eq:Maxwell}
\partial^\nu\partial_\nu A_\mu - \partial_\mu\partial^\nu A_\nu = 0 .
\end{equation}
It only involves the gauge invariant transversal gauge field,
\begin{equation}\label{eq:AT}
A^T_\mu = (P^T)_\mu{}^\nu A_\nu ,
\end{equation}
with projector
\begin{equation}\label{eq:PT}
(P^T)_\mu{}^\nu = \delta_\mu^\nu - \frac{\partial_\mu\partial^\nu}{\partial^2} ,
\qquad \partial^2 = \partial^\mu\partial_\mu .
\end{equation}
The longitudinal field,
\begin{equation}\label{eq:AL}
A^L_\mu = \big(\delta_\mu^\nu - (P^T)_\mu{}^\nu\big)A_\nu
= \frac{\partial_\mu\partial^\nu}{\partial^2}A_\nu ,
\end{equation}
is a gauge degree of freedom and will not be considered further here.
The transversal field obeys
\begin{equation}\label{eq:transversality}
\partial^\mu A^T_\mu = 0 ,
\end{equation}
which reduces the field equation to
\begin{equation}\label{eq:waveAT}
\partial^2 A^T_\mu = 0 .
\end{equation}
Here and in the following repeated indices indicate summation.

Due to the constraint \eqref{eq:transversality} the component $A^T_0$ is not
an independent field. With eq.~\eqref{eq:waveAT} it can be computed from
$A^T_k$ ($k=1\dots3$, $\Delta=\partial^k\partial_k$) as
\begin{equation}\label{eq:A0T}
A^T_0 = \Delta^{-1}\partial_t\,\partial^k A^T_k .
\end{equation}
We therefore need not to include $A^T_0$ and keep as independent fields
$A^T_k$. Using
\begin{equation}\label{eq:pi}
\pi_k = \partial_t A^T_k ,
\end{equation}
the field equation \eqref{eq:waveAT} can be written as coupled first order
field equations
\begin{equation}\label{eq:firstorder}
\partial_t A^T_k = \pi_k , \qquad \partial_t \pi_k = \Delta A^T_k .
\end{equation}

The six fields $A^T_k$, $\pi_k$ are unconstrained. In the form
\eqref{eq:firstorder} these field equations do no longer exhibit the original
gauge- and Lorentz-invariance. On the other hand, the field equations
\eqref{eq:firstorder} are invariant under separate space rotations and
rotations of the components $A^T_k$, $\pi_k$. They take the same form as for
three relativistic massless scalar fields ($\sigma_k = A^T_k$) related by an
``internal'' $SO(3)$-transformation acting on the index $k$. We note that
$A^T_0$ in eq.~\eqref{eq:A0T} remains invariant only under combined space and
internal rotations, while it transforms non-trivially under separate
rotations. This is of no concern as long as we only consider the fields
$A^T_k$, $\pi_k$.

The magnetic and electric fields can be expressed in terms of $A^T_k$ and
$\pi_k$,
\begin{align}\label{eq:BE}
B_k &= \varepsilon_k{}^{ij}\partial_i A^T_j , \nonumber\\
E_k &= -\partial_t A^T_k + \partial_k A^T_0
= -\partial_t\big(A^T_k - \Delta^{-1}\partial_k\partial^l A^T_l\big) .
\end{align}
They are divergence free,
\begin{equation}\label{eq:divfree}
\partial^k B_k = 0 , \qquad \partial^k E_k = 0 ,
\end{equation}
where the second relation employs eq.~\eqref{eq:A0T}, which involves, in turn,
the field eq.~\eqref{eq:waveAT} for $A^T_0$. In the presence of sources the
second eqs.~\eqref{eq:BE} and \eqref{eq:divfree} (for $E_k$) will no longer
hold. Taking a time derivative of eq.~\eqref{eq:BE} yields, with
eq.~\eqref{eq:waveAT}, the second pair of Maxwell's equations,
\begin{equation}\label{eq:Maxwell2}
\partial_t B_k = -\varepsilon_k{}^{ij}\partial_i E_j , \qquad
\partial_t E_k = \varepsilon_k{}^{ij}\partial_i B_j .
\end{equation}

With respect to the combined space and internal rotations we can split
$A^T_k$ into a transversal and longitudinal part
\begin{equation}\label{eq:TTTL}
A^{TT}_k = A^T_k - \Delta^{-1}\partial_k\partial^l A^T_l , \qquad
A^{TL}_k = \Delta^{-1}\partial_k\partial^l A^T_l .
\end{equation}
The longitudinal field $A^{TL}_k$ does not contribute to $B_k$ and $E_k$.
Given magnetic and electric fields are therefore not corresponding to a unique
field $A^T_k$. A shift,
\begin{equation}\label{eq:shift}
A^T_k \to A^T_k + \partial_k\lambda ,
\end{equation}
does not change $B_k$ and $E_k$. In vacuum the field equations
\eqref{eq:waveAT}, \eqref{eq:firstorder} hold separately for $A^{TT}_k$ and
$A^{TL}_k$,
\begin{equation}\label{eq:waveTTTL}
\partial_t^2 A^{TT}_k = \Delta A^{TT}_k , \qquad
\partial_t^2 A^{TL}_k = \Delta A^{TL}_k .
\end{equation}
One may therefore be tempted to omit $A^{TL}_k$. We will not do so here in
view of a later coupling to sources. We formulate our statistical ensemble
for the fields $A^T_k$ and $\pi_k$ obeying the simple linear field equations
\eqref{eq:waveTTTL}. We will recall, however, that photons are expected
to belong to the sector $A^{TT}_k$.

In momentum space the different momentum modes decouple. The field equations
take the simple form
\begin{equation}\label{eq:momentumspace}
\partial_t\sigma_k(\vp) = \pi_k(\vp) , \qquad
\partial_t\pi_k(\vp) = -p^2\sigma_k(\vp) .
\end{equation}
Here $\sigma_k(\vp)$ and $\pi_k(\vp)$ are real functions, related to the real
and imaginary parts of the Fourier transform, e.g.
\begin{equation}\label{eq:Fourier}
\sigma_k(\vx) = \int_p \sigma_k(\vp)\big(\cos(\vp\vx) + \sin(\vp\vx)\big) ,
\end{equation}
normalized with
\begin{equation}\label{eq:normalization}
\int_x \sigma_k(\vx)\tilde\sigma_l(\vx) = \int_p \sigma_k(\vp)\tilde\sigma_l(\vp) .
\end{equation}
The system consists of uncoupled harmonic oscillators, one for each $\vp$
and $k$.

\section{Probabilistic classical fields and quantum field theory subsystem}\label{sec:prob}

Probabilistic classical fields are described by a positive and normalized
probability distribution $w(t,\sigma,\pi) = w\big(t;\sigma_k(\vp),\pi_k(\vp)\big)$.
The probabilistic aspect is realized by specifying an initial probability
distribution at some time $t_{\rm in}$. The time dependence of $w$ is given by
the Liouville equation. We can take over the results of
ref.~\cite{CWDS}, with the only modification that we deal with three
real scalars instead of two, and internal symmetry $SO(3)$ instead of $SO(2)$.

We employ for the time-local probabilistic information the real ``classical
wave function'' $q(t,\sigma,\pi)$~\cite{CWQP1,CWIT,CWQFC,CWQF} which is the square root of the
probability distribution
\begin{equation}\label{eq:wq}
w(t,\sigma,\pi) = q^2(t,\sigma,\pi) .
\end{equation}
(Actually, from the point of view of the overall probability distribution for
events at all times and positions the classical wave function is the most
natural object for encoding the time-local probabilistic
information~\cite{CWPW}.) The time evolution of $q$ is given by the
Liouville equation as well, which can be cast into the form of a Schr\"odinger
equation (with Hamiltonian $H_L$ related to the Liouville operator $\hat L$)
\begin{equation}\label{eq:Schroedinger_q}
i\partial_t q = H_L\, q , \qquad H_L = -i\hat L .
\end{equation}
We can write $H_L$ in terms of hermitian operators
\begin{equation}\label{eq:HL}
H_L = \int_p\sum_k\Big(\hat\pi_k(\vp)\hat\gamma_k(\vp)
+ p^2\hat\sigma_k(\vp)\hat\zeta_k(\vp)\Big) ,
\end{equation}
where in the present basis
\begin{align}\label{eq:operators}
\hat\sigma_k(\vp) &= \sigma_k(\vp) , \qquad \hat\pi_k(\vp) = \pi_k(\vp) ,
\nonumber\\
\hat\gamma_k(\vp) &= -i\frac{\partial}{\partial\sigma_k(\vp)} , \qquad
\hat\zeta_k(\vp) = i\frac{\partial}{\partial\pi_k(\vp)} .
\end{align}
For discrete momenta $\int_p$ denotes the sum over momenta, which corresponds
for continuous $\vp$ to $\int_p = \Omega(2\pi)^{-3}\int d^3p$, with $\Omega$ the
volume of space.

For linear field equations $H_L$ can be decomposed by a
linear operator transformation,
\begin{align}\label{eq:phichi}
\hat\varphi_k(\vp) &= \hat\sigma_k(\vp) + \tfrac12\hat\zeta_k(\vp) , \qquad
\hat\chi_k(\vp) = \hat\sigma_k(\vp) - \tfrac12\hat\zeta_k(\vp) , \nonumber\\
\hat\eta_k(\vp) &= \hat\pi_k(\vp) + \tfrac12\hat\gamma_k(\vp) , \qquad
\hat\beta_k(\vp) = \hat\pi_k(\vp) - \tfrac12\hat\gamma_k(\vp) ,
\end{align}
which yields
\begin{equation}\label{eq:HLsplit}
H_L = H^{(q)} - H^{(m)} ,
\end{equation}
where
\begin{align}\label{eq:Hq_Hm}
H^{(q)} &= \frac12\int_p\sum_k\Big(\hat\eta_k(\vp)\hat\eta_k(\vp)
+ p^2\hat\varphi_k(\vp)\hat\varphi_k(\vp)\Big) , \nonumber\\
H^{(m)} &= \frac12\int_p\sum_k\Big(\hat\beta_k(\vp)\hat\beta_k(\vp)
+ p^2\hat\chi_k(\vp)\hat\chi_k(\vp)\Big) .
\end{align}
For these operators the non-vanishing commutation relations read
\begin{align}\label{eq:commutators}
\big[\hat\varphi_k(\vp),\hat\eta_l(\vp\,')\big] &= i\,\delta_{kl}\,\delta(\vp-\vp\,') ,
\nonumber\\
\big[\hat\chi_k(\vp),\hat\beta_l(\vp\,')\big] &= -i\,\delta_{kl}\,\delta(\vp-\vp\,') .
\end{align}

As familiar in quantum mechanics we can introduce time-dependent Heisenberg
operators. Their time evolution obeys
\begin{align}\label{eq:Heisenberg}
\partial_t\hat\varphi_k(\vp) &= i\big[H_L,\hat\varphi_k(\vp)\big]
= i\big[H^{(q)},\hat\varphi_k(\vp)\big] = \hat\eta_k(\vp) , \nonumber\\
\partial_t\hat\eta_k(\vp) &= i\big[H_L,\hat\eta_k(\vp)\big]
= i\big[H^{(q)},\hat\eta_k(\vp)\big] = -p^2\hat\varphi_k(\vp) .
\end{align}
These are the evolution equations for field operators in a quantum field
theory for the electromagnetic field. They are combined with the canonical
commutation relation \eqref{eq:commutators}. We can associate $\hat\varphi_k$
with the operator for the quantum field $A^{T(q)}_k$ and $\hat\eta_k$ with
$\partial_t A^{T(q)}_k$. In contrast, $\hat\sigma_k$ and $\hat\pi_k$ are the
operators associated to the classical fields $A^{T(\mathrm{cl})}_k$ and
$\partial_t A^{T(\mathrm{cl})}_k$. The quantum and classical fields differ. For
example, the operators $\hat\sigma_k$ and $\hat\pi_k$ commute, in distinction
to $\hat\varphi_k$ and $\hat\eta_k$.

The operator relations \eqref{eq:HLsplit}--\eqref{eq:Heisenberg} hold in an
arbitrary basis. We can bring our setting even closer to the standard quantum
field theory formulation by making a change of basis. A functional Fourier
transform maps the real wave function $q$ to a complex wave function
$\tilde\psi$,
\begin{align}\label{eq:FFT}
\tilde\psi(\varphi,\chi) = \int\mathcal{D}\pi\,
&\exp\Big\{i\int_p\sum_k\big(\varphi_k(\vp)-\chi_k(\vp)\big)\pi_k(\vp)\Big\}
\nonumber\\
&\times q\Big(\frac{\varphi+\chi}{2},\pi\Big) .
\end{align}
Since $q$ is real, the complex wave function $\tilde\psi$ obeys the constraint
\begin{equation}\label{eq:psiconstraint}
\tilde\psi^*(\varphi,\chi) = \tilde\psi(\chi,\varphi) .
\end{equation}
In this basis the operators are given by
\begin{equation}\label{eq:opsphi}
\hat\varphi_k(\vp) = \varphi_k(\vp) , \qquad
\hat\eta_k(\vp) = -i\frac{\partial}{\partial\varphi_k(\vp)} .
\end{equation}
In the opposite direction $q(\sigma,\pi)$ is the Wigner transform~\cite{WIG}
of $\tilde\psi(\varphi,\chi)$.

In this paper we investigate pure quantum states where
\begin{equation}\label{eq:purestate}
\tilde\psi(\varphi,\chi) = \psi(\varphi)\,\psi^*(\chi) .
\end{equation}
The wave function $\psi(\varphi)$ for the quantum fields $\varphi$ is an
arbitrary normalized complex function. Its evolution obeys the Schr\"odinger
equation
\begin{align}\label{eq:Schroedinger_psi}
i\partial_t\psi(\varphi) &= H^{(q)}\psi(\varphi) \nonumber\\
&= \frac12\int_p\sum_k\Big(-\frac{\partial^2}{\partial\varphi_k(\vp)^2}
+ p^2\varphi_k^2(\vp)\Big)\psi(\varphi) .
\end{align}
We can now construct the wave function $\psi(\varphi)$ for an entangled
two-photon state of the quantum field theory for electromagnetic fields. Then
eq.~\eqref{eq:purestate} yields $\tilde\psi(\varphi,\chi)$, and inverting the
functional Fourier transform \eqref{eq:FFT} one obtains the real wave function
$q(\sigma,\pi)$. From there one obtains the classical probability distribution
$w(\sigma,\pi)$. Our construction guarantees that every solution of the
Schr\"odinger eq.~\eqref{eq:Schroedinger_psi} corresponds to a solution of the
Liouville equation for $w$. To every pure state wave function $\psi(\varphi)$
one can associate a corresponding classical probability distribution
$w(\sigma,\pi)$.

\section{Photon states}\label{sec:photons}

The procedure for constructing wave functions for multi-photon states is
standard. For every momentum mode $\vp$ we introduce annihilation and creation
operators
\begin{align}\label{eq:aadagger}
a_k(\vp) &= \sqrt{\frac{p}{2}}\,\hat\varphi_k(\vp)
+ \frac{i}{\sqrt{2p}}\,\hat\eta_k(\vp) , \nonumber\\
a^\dagger_k(\vp) &= \sqrt{\frac{p}{2}}\,\hat\varphi_k(\vp)
- \frac{i}{\sqrt{2p}}\,\hat\eta_k(\vp) ,
\end{align}
obeying the commutation relation
\begin{equation}\label{eq:aacomm}
\big[a_k(\vp),a^\dagger_l(\vp)\big] = \delta_{kl} .
\end{equation}
The quantum Hamiltonian $H^{(q)}$ is a sum over photon occupation numbers
\begin{equation}\label{eq:Hqn}
H^{(q)} = \int_p\sum_k p\Big(\hat n_k(\vp) + \frac12\Big) ,
\end{equation}
\begin{equation}\label{eq:n}
\hat n_k(\vp) = a^\dagger_k(\vp)\,a_k(\vp) .
\end{equation}

The quantum vacuum wave function $\psi_0$ is defined by
\begin{equation}\label{eq:vacuum}
a_k(\vp)\,\psi_0 = 0 .
\end{equation}
The function $\psi_0(\varphi)$ is Gaussian
\begin{equation}\label{eq:psi0}
\psi_0(\varphi) = \mathcal{N}_\psi\exp\Big\{-\frac12\sum_k\int_p
p\,\varphi_k(\vp)\varphi_k(\vp)\Big\} .
\end{equation}
This vacuum has the ground state energy
\begin{equation}\label{eq:E0}
E_0 = \frac32\int_p\sqrt{p^2} ,
\end{equation}
which is ultraviolet divergent if the $p$-integration extends to infinitely
large $|p|$. However, a phase factor $\exp(-iE_0t)$ in $\psi(\varphi)$ is
cancelled by a similar factor for $\psi^*(\chi)$ in eq.~\eqref{eq:purestate}.
Thus $E_0$ does not appear in $\tilde\psi$, $q$ and $w$. The quantum energy of photon states subtracts the unobservable ground state
energy
\begin{equation}\label{eq:Hqtilde}
\tilde H^{(q)} = H^{(q)} - E_0 = \int_p\sum_k p\,\hat n_k(\vp) .
\end{equation}
It is positive.

The classical probability distribution for the quantum vacuum is a Gaussian
with finite width,
\begin{align}\label{eq:wvacuum}
w(\sigma,\pi) = \mathcal{N}_0^2\exp\Big\{&-2\sum_k\int_p\Big(
p\,\sigma_k(\vp)\sigma_k(\vp)\nonumber\\
&+ \frac1p\,\pi_k(\vp)\pi_k(\vp)\Big)\Big\} ,
\end{align}
with $\mathcal{N}_0$ a normalization factor. This differs from the ``classical
vacuum'' for which $w$ is a $\delta$-distribution singling out zero fields.

A basis for the quantum wave function $\psi(\varphi)$ can be constructed from
the photon occupation numbers. A basis state specifies for each $\vp$ and $k$ the
number $\hat n_k(\vp)$ of photons. For the free electromagnetic field each
occupation number $\hat n_k(\vp)$ corresponds to a conserved quantity. We can
therefore define subsystems for a given total photon number
$N = \sum_k\int_p n_k(\vp)$. These subsystems are closed under the evolution.

A one-photon state with a sharp momentum $\vp$ obtains from the vacuum state
by applying the creation operator $a^\dagger_k(\vp)$,
\begin{equation}\label{eq:onephoton}
\psi_1(\varphi) = \sum_k c_k\, a^\dagger_k(\vp)\,\psi_0(\varphi) .
\end{equation}
It is an eigenstate of $\hat n(\vp) = \sum_l \hat n_l(\vp)$,
\begin{equation}\label{eq:n1}
\hat n(\vp)\,\psi_1(\varphi) = \psi_1(\varphi) .
\end{equation}
Physical photons are transversal. For $\vp = (0,0,p_3)$ a transversal photon
state reads
\begin{equation}\label{eq:transphoton}
\psi_1(\varphi) = \big(c_1a^\dagger_1(\vp) + c_2a^\dagger_2(\vp)\big)\psi_0 .
\end{equation}
The general case can be obtained by rotation. A physical photon state obeys
\begin{equation}\label{eq:transcond}
\sum_k c_k\,p_k = 0 .
\end{equation}
The two independent $c_k$ indicate the polarization of the photon. For our
example $\vp=(0,0,p_3)$ the choice $c_2=0$ corresponds to transversal
polarization in the one-direction. Instead of a fixed $\vp$ one can extend the
discussion to wave packets with average momentum $\vp$.

Let us next construct a state of two (transversal) photons with opposite
momentum,
\begin{align}\label{eq:twophoton}
\psi_2(\varphi) &= \sum_{k,l} c_{kl}\,a^\dagger_k(\vp)\,a^\dagger_l(-\vp)\,\psi_0 ,
\nonumber\\
\sum_l c_{kl}\,p_l &= \sum_l c_{lk}\,p_l = 0 .
\end{align}
For $\vp=(0,0,p_3)$ this depends on the four coefficients $c_{11}$, $c_{12}$,
$c_{21}$, $c_{22}$. In general, the two photons are correlated and we deal
with an entangled state. A direct product state obtains only if $c_{kl}$
factorizes, $c_{kl} = a_k b_l$. For example the choice $c_{12}=c_{21}=0$, $c_{22}=c_{11}$
corresponds to two photons with correlated polarization. If photon one is
polarized in the one-direction, also photon two is polarized in the
one-direction. The same holds for the polarization in the two-direction.

We can explicitly indicate the classical probability distribution
$w_2(\sigma,\pi)$ which corresponds to a two-photon state. For the example
\begin{align}\label{eq:psi2ex}
\psi_2(\varphi) &= \frac{1}{\sqrt2}\Big(a^\dagger_1(\vp)a^\dagger_1(-\vp)
+ a^\dagger_2(\vp)a^\dagger_2(-\vp)\Big)\psi_0 \nonumber\\
&= \sqrt2\,p\,\Big(\varphi_1(\vp)\varphi_1(-\vp)
+ \varphi_2(\vp)\varphi_2(-\vp)\Big)\psi_0(\varphi)
\end{align}
we employ eq.~\eqref{eq:psi0} and the explicit form of the creation operator
\eqref{eq:aadagger}. For a pure state wave function \eqref{eq:purestate} this
yields $\tilde\psi_2(\varphi,\chi)$. The real classical wave function
$q_2(\sigma,\pi)$ is the Wigner transform of $\tilde\psi_2(\varphi,\chi)$,
\begin{align}\label{eq:q2}
q_2(\sigma,\pi) = \Big\{&\frac12\Big[\big(2e_1(\vp)-1\big)\big(2e_1(-\vp)-1\big)
\nonumber\\
&+\big(2e_2(\vp)-1\big)\big(2e_2(-\vp)-1\big)\Big]\nonumber\\
&+4\Big[s(\vp)s(-\vp)-l(\vp)l(-\vp)\Big]\Big\}\,q_0(\sigma,\pi) .
\end{align}
It is real, but not positive. The probability distribution
$w_2(\sigma,\pi)=q_2^2(\sigma,\pi)$ is then positive and normalized. In
eq.~\eqref{eq:q2} we use the abbreviations
\begin{align}\label{eq:q2def}
e_k(\vp) &= p\,\sigma_k^2(\vp) + \frac1p\,\pi_k^2(\vp) , \nonumber\\
s(\vp) &= p\,\sigma_1(\vp)\sigma_2(\vp) + \frac1p\,\pi_1(\vp)\pi_2(\vp) ,
\nonumber\\
l(\vp) &= \sigma_1(\vp)\pi_2(\vp) - \sigma_2(\vp)\pi_1(\vp) ,
\end{align}
and $q_0$ is the classical wave function of the vacuum,
\begin{align}\label{eq:q0}
q_0(\sigma,\pi) = \mathcal{N}_0\exp\Big\{&-\sum_k\int_p\Big(p\,\sigma_k(\vp)
\sigma_k(\vp)\nonumber\\
&+ \frac1p\,\pi_k(\vp)\pi_k(\vp)\Big)\Big\} ,
\end{align}
such that $q_0^2$ is the vacuum distribution \eqref{eq:wvacuum}.

We emphasize that $w_2(\sigma,\pi)=q_2^2(\sigma,\pi)$ is a positive and
normalized classical probability distribution for the classical fields
$\sigma_k(\vp)$ and $\pi_k(\vp)$. Its time evolution is given by the
Liouville equation for the classical field equations \eqref{eq:momentumspace}.
The entangled two-photon state is therefore a state of a classical
probabilistic system, without any negative or complex probabilities.

\begin{figure*}[t]
\centering
\includegraphics[width=\textwidth]{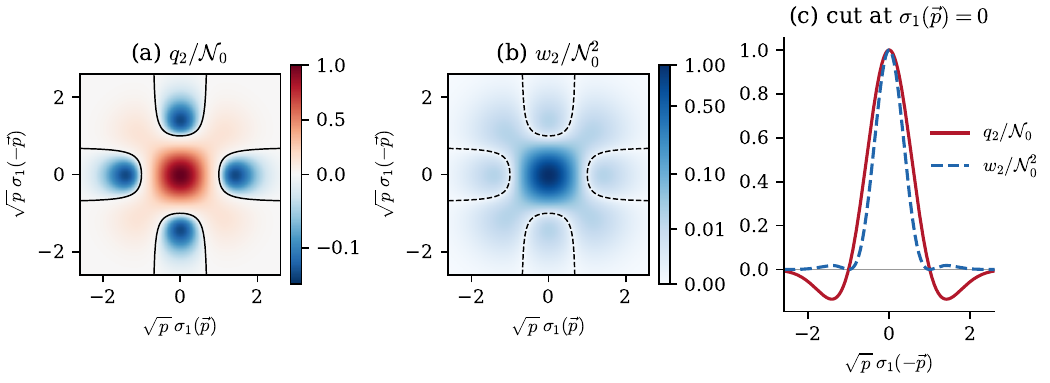}
\caption{Classical wave function $q_2$ and probability distribution
$w_2=q_2^2$ for the entangled two-photon state \eqref{eq:psi2ex}, on the
cut where all fields except $\sigma_1(\vp)$ and $\sigma_1(-\vp)$ vanish,
cf.\ eq.~\eqref{eq:q2cut}. (a) $q_2/\mathcal N_0$; solid lines mark
$q_2=0$, and the blue regions are negative. (b) $w_2/\mathcal N_0^2$ on a
compressed color scale; it is positive everywhere and vanishes only on the
dashed lines. (c) Cut along $\sigma_1(\vp)=0$.}
\label{fig:q2}
\end{figure*}

In Fig.~\ref{fig:q2} we show $q_2$ and $w_2$ for a two-dimensional cut
through field space, where all fields except $\sigma_1(\vp)$ and
$\sigma_1(-\vp)$ are set to zero. With $x=\sqrt p\,\sigma_1(\vp)$ and
$y=\sqrt p\,\sigma_1(-\vp)$ eq.~\eqref{eq:q2} reduces on this cut to
\begin{equation}\label{eq:q2cut}
q_2 = \frac{\mathcal N_0}{2}\Big[\big(2x^2-1\big)\big(2y^2-1\big)+1\Big]
e^{-x^2-y^2} .
\end{equation}
The classical wave function $q_2$ takes negative values in four regions,
bounded by lines on which $q_2$ vanishes. In contrast, $w_2=q_2^2$ is
positive everywhere. Up to normalization, $q_2$ is the Wigner function of
the two-photon state. Bell has discussed Wigner functions as probability
distributions for local hidden variables~\cite{BELL86}. Their negativity for
entangled states is an obstacle to such an interpretation. In our setting
the negative values of $q_2$ are no obstacle, since the probability
distribution is $w_2=q_2^2$.

\section{Polarization measurements}
\label{sec:polmeas}

For any real measurement one needs a rule which relates quantities in the
mathematical formalism to the outcomes of the measurement apparatus. This is,
in general, a rather complex issue. One typically considers observables for
which measurements produce a probability distribution for the possible
outcomes. For ideal measurements one assumes that the expectation value
constructed from the measured probabilities corresponds to the expectation
value in the mathematical formalism. In our quantum formalism for classical
probabilistic theories the expectation values of linear observables are
computed from the ``classical'' wave function $q$ by associating an operator
to the observable and evaluating the expectation value as a bilinear in $q$,
similar to the usual quantum rule. This rule holds directly for all classical
observables which are represented as functions of $\sigma$ and $\pi$. It is
extended to statistical observables for which the operators involve
derivatives with respect to $\sigma$ and $\pi$.

\subsection{Polarization detector}

For photons with a given momentum $\vp$ we have seen that there are two
transversal polarization directions. For simplicity we consider first a state
with a single photon with $\vp=(0,0,p_3)$. We then have two different
``photon types''. One with polarization in the one-direction, with associated
photon number $\hat n_1$, and the other in the orthogonal two-direction, with
photon number $\hat n_2$. For a one photon state $\psi_1$ one has
\begin{equation}\label{eq:PM1}
(\hat n_1+\hat n_2)\,\psi_1 = \psi_1 .
\end{equation}
A key feature of polarization measurements is the assumption that one can
measure $\hat n_1$ separately. The measurement apparatus corresponds to an
observable which takes the value one if a photon of type one is present, and
zero if no photon of type one is present. It is typically realized by a
polarization filter which lets all photons polarized in the one-direction
pass, and rejects all photons polarized in the two-direction. Behind the
polarization filter one places a photon detector. The possible outcomes of
the measurement, $1$ or $0$, correspond to the eigenvalues of the operator
$\hat n_1$.

For the particular one-photon state
\begin{equation}\label{eq:PM2}
\psi_{11} = a^\dagger_1\psi_0 , \qquad \hat n_1\psi_{11} = \psi_{11} ,
\end{equation}
the probability to find a photon of type one equals one. Correspondingly,
for the orthogonal one-photon state,
\begin{equation}\label{eq:PM3}
\psi_{12} = a^\dagger_2\psi_0 , \qquad \hat n_1\psi_{12} = 0 ,
\end{equation}
the probability to find a photon of type one equals zero. This follows from
the fact that the total probability to find a photon in a one-photon state
equals one, and the probability to find a photon of type two also equals
one. We can define the mathematical probability $w_1$ to find a photon of
type one by
\begin{equation}\label{eq:PM4}
w_1 = \langle n_1\rangle = \psi_1^\dagger\,\hat n_1\,\psi_1 ,
\end{equation}
and similarly
\begin{equation}\label{eq:PM5}
w_2 = \langle n_2\rangle = \psi_1^\dagger\,\hat n_2\,\psi_1 ,
\end{equation}
with $w_1+w_2=1$, $w_{1,2}\ge0$. For a general one-photon state one finds
\begin{align}\label{eq:PM6}
\psi_1 &= c_1\psi_{11} + c_2\psi_{12} , \nonumber\\
\langle n_1\rangle &= |c_1|^2 , \qquad \langle n_2\rangle = |c_2|^2 .
\end{align}
If we use a polarization basis
\begin{equation}\label{eq:PM7}
\hat\psi_1 = \begin{pmatrix} c_1\\ c_2\end{pmatrix} ,
\end{equation}
the probabilities
\begin{equation}\label{eq:PM8}
w_1 = |c_1|^2 , \qquad w_2 = |c_2|^2 ,
\end{equation}
are Born's rule for the mathematical probabilities defined by
eqs.~\eqref{eq:PM4}, \eqref{eq:PM5}. It follows from the definition of
expectation values for linear observables.

For an ideal polarization detector one assumes that the probability that it
fires equals the mathematical probability $w_1=\langle n_1\rangle$. This is
Born's rule for an ideal measurement. The operator $\hat n_1$ fulfills all
requirements for a polarization detector. It is a very natural choice. If at
all a linear observable can represent a polarization detector we see no
reasonable other choice. All this is very similar to the setting in quantum
mechanics where one has to make similar assumptions. The only difference is
that $\hat n_1$ is here an operator acting on the classical wave function $q$
in a classical probabilistic setting. The operator $\hat n_1$ is the same
as the one used for photon detectors in quantum field theory. Our assumption
is therefore not stronger than the standard one in quantum field theory,
where the association of an observable to a given measurement apparatus is
equally an assumption. It is a central assumption of this
paper that a polarization detector in the one-direction of the filter is
represented by the observable $n_1$, with probability to detect a photon
given by $\langle n_1\rangle$. Starting from there, more complex polarization
measurements will finally lead to a violation of Bell's inequalities
for pairs of correlated photons. This outcome is based on the simple concept
of the observable $n_1$.

It is convenient to define a polarization observable as
\begin{equation}\label{eq:PM9}
\hat P(\alpha=0) = \hat n_1 - \hat n_2 .
\end{equation}
For one-photon states it has the possible measurement value $+1$ if a photon
is detected after the polarization filter, and $-1$ if no photon is detected.
With $\hat n_2 = 1-\hat n_1$ we can also write $\hat P(\alpha=0)=2\hat n_1-1$. The
angle $\alpha=0$ denotes the orientation of the polarization filter in the
one-direction. We may turn the filter by an angle $\pi/2$ around the
three-axis, such that photons of type two pass,
\begin{equation}\label{eq:PM10}
\hat P(\alpha=\pi/2) = \hat n_2 - \hat n_1 = -\hat P(\alpha=0) .
\end{equation}

\subsection{Rotations}

A polarization detector with arbitrary angle $\alpha$ of the polarization
filter can be obtained from $P(\alpha=0)$ by rotation. The corresponding
operator is given by
\begin{equation}\label{eq:PM11}
\hat P(\alpha) = \hat n_\alpha - \hat n_{\alpha T} ,
\end{equation}
with
\begin{equation}\label{eq:PM12}
\hat n_\alpha = a^\dagger_\alpha a_\alpha , \qquad
\hat n_{\alpha T} = a^\dagger_{\alpha T}a_{\alpha T} ,
\end{equation}
and
\begin{align}\label{eq:PM13}
a_\alpha &= \cos\alpha\,a_1 + \sin\alpha\,a_2 , \nonumber\\
a_{\alpha T} &= \cos\alpha\,a_2 - \sin\alpha\,a_1 .
\end{align}
This yields
\begin{equation}\label{eq:PM14}
\hat P(\alpha) = \cos(2\alpha)\,(\hat n_1-\hat n_2)
+ \sin(2\alpha)\,(a^\dagger_2a_1 + a^\dagger_1a_2) .
\end{equation}

For two general angles $\alpha$ and $\gamma$ the polarization operators do
not commute
\begin{align}\label{eq:PM15}
\big[\hat P(\alpha),\hat P(\gamma)\big]
&= B(\alpha,\gamma)\big[(\hat n_1-\hat n_2),(a^\dagger_2a_1+a^\dagger_1a_2)\big]
\nonumber\\
&= 2B(\alpha,\gamma)\,(a^\dagger_1a_2 - a^\dagger_2a_1) ,
\end{align}
where
\begin{equation}\label{eq:PM16}
\begin{split}
B(\alpha,\gamma) &= \sin(2\gamma)\cos(2\alpha) - \sin(2\alpha)\cos(2\gamma)\\
&= \sin\big(2(\gamma-\alpha)\big) .
\end{split}
\end{equation}
This non-commutativity of polarization operators is a crucial ingredient for
the violation of Bell's inequalities which will be discussed in the next
section.

We need to show how the polarization observable
\eqref{eq:PM11}--\eqref{eq:PM14} follows from the polarization observable
$P(\alpha=0)$ in eq.~\eqref{eq:PM9}, and therefore finally from the detector
observable $n_1$ for the presence of a photon of type one. This can be
achieved by rotation. Our setting for one-photon states can be rotated around
the three-axis. Rotation symmetry of the dynamics implies that the
expectation value of the polarization observable does not change if one
rotates simultaneously the photon state and the polarization operator. In
turn, a rotation of the polarization observable for a fixed photon state is
equivalent to the rotation of the photon state in the opposite direction,
keeping the polarization observable fixed.

Our formulation based on the classical wave function $q$ permits a
straightforward description of the effects of rotations on the photon
states. We have seen that the classical field equations are invariant under
simultaneous ``internal'' rotations of the field components $\sigma_k$ and
$\pi_k$. The action of a rotation with infinitesimal angle $\delta_k$ on the
wave function obeys
\begin{align}\label{eq:R1}
\delta q(\sigma,\pi) ={}& \sum_l\int_p\Big(\frac{\partial q}{\partial\sigma_l(\vp)}\,
\delta\sigma_l(\vp) + \frac{\partial q}{\partial\pi_l(\vp)}\,\delta\pi_l(\vp)\Big)
\nonumber\\
={}& \sum_{l,k,m}\int_p\Big(\frac{\partial q}{\partial\sigma_l(\vp)}\,\sigma_m(\vp)
+ \frac{\partial q}{\partial\pi_l(\vp)}\,\pi_m(\vp)\Big)\nonumber\\
&\times\varepsilon_{lkm}\,\delta_k .
\end{align}
It involves the generator of internal rotations
\begin{align}\label{eq:R2}
\delta q &= i\sum_k\delta_k\,\hat L_k\,q , \nonumber\\
\hat L_k &= -i\sum_{m,l}\varepsilon_{kml}\int_p\Big(\sigma_m(\vp)
\frac{\partial}{\partial\sigma_l(\vp)}
+ \pi_m(\vp)\frac{\partial}{\partial\pi_l(\vp)}\Big) .
\end{align}
With
\begin{equation}\label{eq:R3}
\big[H_L,\hat L_k\big] = 0 ,
\end{equation}
we can associate arbitrary functions of the operators $\hat L_k$ with
conserved statistical observables. These statistical observables
characterize the properties of the probabilistic information under internal
rotations. For example, an eigenstate with $\hat L_k q=0$ implies that $q$ is
invariant under rotations around the $k$-axis.

The action of infinitesimal rotations carries over to $\tilde\psi$
\begin{align}\label{eq:R4}
\delta\tilde\psi &= i\sum_k\delta_k\,\hat L_k\,\tilde\psi , \nonumber\\
\hat L_k &= -i\sum_{l,m}\varepsilon_{klm}\int_p\Big(\varphi_l(\vp)
\frac{\partial}{\partial\varphi_m(\vp)}
+ \chi_l(\vp)\frac{\partial}{\partial\chi_m(\vp)}\Big) ,
\end{align}
as well as to the complex wave function $\psi(\varphi)$ in case of a pure
state
\begin{align}\label{eq:R5}
\delta\psi &= i\sum_k\delta_k\,\hat L^{(q)}_k\,\psi , \nonumber\\
\hat L^{(q)}_k &= -i\sum_{l,m}\varepsilon_{klm}\int_p\varphi_l(\vp)
\frac{\partial}{\partial\varphi_m(\vp)} .
\end{align}
We observe that $\hat L^{(q)}_k$ is conserved separately,
\begin{equation}\label{eq:R6}
\big[\hat L^{(q)}_k,H^{(q)}\big] = 0 , \qquad \big[\hat L^{(q)}_k,H_L\big] = 0 .
\end{equation}
The vacuum state is rotation invariant
\begin{equation}\label{eq:R7}
\hat L^{(q)}_k\,\psi_0 = 0 .
\end{equation}
This implies for the rotation of the one-photon state \eqref{eq:onephoton}
\begin{align}\label{eq:R8}
\delta\psi_1 &= \delta\Big(\sum_k c_k\,a^\dagger_k(\vp)\,\psi_0(\varphi)\Big)
\nonumber\\
&= i\sum_{l,k}\delta_l\,c_k\big[\hat L^{(q)}_l,a^\dagger_k(\vp)\big]\psi_0(\varphi)
\nonumber\\
&= -\sum_{l,k,m}\delta_l\,c_k\,\varepsilon_{lkm}\,a^\dagger_m(\vp)\,\psi_0(\varphi) .
\end{align}
The last line follows from the explicit definition \eqref{eq:aadagger} of
$a^\dagger_k(\vp)$. We conclude that a rotation of a one-photon state
corresponds to a rotation of the creation operators $a^\dagger_k(\vp)$, which
transform as vectors. This extends to multi-photon states.

A simultaneous rotation of the photon states and the operator for the
polarization detector corresponds to a change of basis. This results in a
simultaneous rotation of all creation and annihilation operators. In
consequence, the rotation of the photon state \eqref{eq:R8} can be
transferred to a rotation of the annihilation and creation operators forming
the polarization operator, while the photon state remains invariant.
Applying a rotation in the plane orthogonal to $\vp$ results in the operator
$\hat P(\alpha)$ for a polarization detector at an arbitrary angle $\alpha$, as
given by eqs.~\eqref{eq:PM11}--\eqref{eq:PM14}. This form is not an
``ad hoc'' ansatz. We have shown that it follows from eq.~\eqref{eq:PM9}. The
chain of argument is very similar to what one does in a quantum field theory.
We have demonstrated here that it applies to a classical field theory as
well. The formulation in terms of the classical wave function $q$ has been
very useful for this purpose.

\subsection{Photons as qubits}

The one-photon state \eqref{eq:transphoton} is a two-level quantum state. It
can be associated with a qubit. The general wave function can be
characterized by the two-component complex wave function
\begin{equation}\label{eq:PQ1}
c = \begin{pmatrix}c_1\\ c_2\end{pmatrix} , \qquad c^\dagger c = 1 .
\end{equation}
We can express the action of the polarization operator $\hat P(\alpha)$ in this
language
\begin{equation}\label{eq:PQ2}
\hat P(\alpha)\begin{pmatrix}c_1\\ c_2\end{pmatrix}
= \begin{pmatrix}c'_1\\ c'_2\end{pmatrix} ,
\end{equation}
where
\begin{equation}\label{eq:PQ3}
\sum_k \hat P(\alpha)\,c_k\,a^\dagger_k\psi_0 = \sum_k c'_k\,a^\dagger_k\psi_0 .
\end{equation}
In terms of the Pauli-matrices $\tau_k$ this amounts to
\begin{equation}\label{eq:PQ4}
\hat P(\alpha) = \cos(2\alpha)\,\tau_3 + \sin(2\alpha)\,\tau_1 .
\end{equation}

Two entangled photons correspond to two entangled qubits. This makes it
plausible that correlations between two photons can violate Bell's
inequalities.

\section{Polarization observables and Bell inequalities}
\label{sec:Bell}

The entangled two-photon states \eqref{eq:twophoton} are the quantum states
for which violations of Bell's inequalities~\cite{BEL1,CHSH} have been
established experimentally, starting with atomic
cascades~\cite{FRCL,ASP1,ASP2,ASP3} and continuing with parametric
down-conversion~\cite{KWI,WEI}, up to the loophole-free
tests~\cite{HEN,GIU,SHA}. In this section we employ the polarization
observables of sect.~\ref{sec:polmeas} for the two photons and compute their
correlations for an arbitrary two-photon state. We show explicitly that for every entangled state the
Clauser--Horne--Shimony--Holt (CHSH) inequality~\cite{CHSH} is violated. Since
each of these quantum states is associated to a positive classical probability
distribution $w(\sigma,\pi)$ for the classical fields, the probabilistic
classical field theory of sects.~\ref{sec:field} and \ref{sec:prob} contains states with this property.
We discuss at the end of this section why this does not contradict Bell's
theorem.

\subsection{Two-photon subspace}

For definiteness we work with discrete momenta, for which $\int_p$ is a sum
and $\delta(\vp-\vp\,')$ in eq.~\eqref{eq:commutators} is replaced by the
Kronecker symbol $\delta_{\vp\vp\,'}$. (For continuous momenta the norm of a
state with sharp momenta involves factors $\delta(0)$, and one has to work
with wave packets. The structure of the polarization correlations is not
affected.) The annihilation and creation operators \eqref{eq:aadagger} then
obey
\begin{equation}\label{eq:aacommdiscrete}
\big[a_k(\vp),a^\dagger_l(\vp\,')\big] = \delta_{kl}\,\delta_{\vp\vp\,'} .
\end{equation}

We take $\vp=(0,0,p_3)$, $p_3>0$. The transversality conditions in
eq.~\eqref{eq:twophoton} imply $c_{3l}=c_{k3}=0$, such that the two-photon
states with momenta $\vp$ and $-\vp$ form a four-dimensional space with basis
\begin{equation}\label{eq:basis2}
|kl\rangle = a^\dagger_k(\vp)\,a^\dagger_l(-\vp)\,\psi_0 , \qquad k,l\in\{1,2\} .
\end{equation}
For $\vp\neq-\vp$ these basis states are orthonormal,
$\langle kl|mn\rangle=\delta_{km}\delta_{ln}$. A general two-photon state,
\begin{equation}\label{eq:psi2C}
\psi_2 = \sum_{k,l=1}^2 c_{kl}\,|kl\rangle ,
\end{equation}
is characterized by the complex $2\times2$ matrix $C=(c_{kl})$. We normalize
\begin{equation}\label{eq:normC}
\langle\psi_2|\psi_2\rangle = \sum_{k,l}|c_{kl}|^2 = \mathrm{tr}\big(C^\dagger C\big) = 1 .
\end{equation}
The first index of $c_{kl}$ refers to the polarization of the photon with
momentum $\vp$ (photon $A$), the second one to the photon with momentum $-\vp$
(photon $B$). In this language eq.~\eqref{eq:psi2C} is the familiar state of
two qubits, $\psi_2=\sum_{kl}c_{kl}|k\rangle_A\otimes|l\rangle_B$. It is a
product state if and only if $C$ has rank one, i.e.\ $\det C=0$. A convenient
measure of entanglement is the concurrence~\cite{WOO}
\begin{equation}\label{eq:concurrence}
\mathcal{C} = 2\,|\det C| = 2\,|c_{11}c_{22}-c_{12}c_{21}| ,
\qquad 0\le\mathcal{C}\le1 .
\end{equation}
Maximally entangled states have $\mathcal{C}=1$. Two important examples are
\begin{equation}\label{eq:examples}
C_+ = \frac{1}{\sqrt2}\begin{pmatrix}1&0\\0&1\end{pmatrix} , \qquad
C_- = \frac{1}{\sqrt2}\begin{pmatrix}0&1\\-1&0\end{pmatrix} .
\end{equation}
The state $C_+$, with $c_{12}=c_{21}=0$, is the state with correlated
polarization discussed at the end of sect.~\ref{sec:photons}. It is the state produced in
the $J=0\to1\to0$ atomic cascade used in the experiments of
refs.~\cite{FRCL,ASP1,ASP2,ASP3}. The state $C_-$, with $c_{kl}=\varepsilon_{kl3}/\sqrt2$,
describes two photons with orthogonal linear polarization. It is the
two-photon state from the decay of a pseudoscalar, as for
$\pi^0\to\gamma\gamma$ or the annihilation of para-positronium, where the
decay amplitude is proportional to
$(\vec\varepsilon_A\times\vec\varepsilon_B)\cdot\vp$~\cite{YAN}. The
orthogonal polarization of the two photons has been used to determine the
parity of the neutral pion~\cite{PLA}, and the polarization correlation of
annihilation photons has been observed by Compton
polarimetry~\cite{WUSH,KUW}. For photons in the MeV range no ideal
polarizers exist, however, and Compton polarimetry alone does not permit a
test of Bell's inequalities without additional
assumptions~\cite{KUW,HIMO}. Experimentally, the state $C_+$ is therefore
the relevant one.

\subsection{Polarization operators for two photons}

For each of the two photons we use the polarization operator of
sect.~\ref{sec:polmeas}, with angles measured with respect to the same fixed
axes $1$ and $2$ for both photons. For the photon with momentum $\vp$,
eqs.~\eqref{eq:PM11}--\eqref{eq:PM14} read
\begin{equation}\label{eq:PalphaStokes}
\hat P_\alpha(\vp) = \cos2\alpha\,\hat S_3(\vp) + \sin2\alpha\,\hat S_1(\vp) ,
\end{equation}
where we have introduced the Stokes operators
\begin{align}\label{eq:Stokes}
\hat S_1(\vp) &= a^\dagger_1(\vp)a_2(\vp) + a^\dagger_2(\vp)a_1(\vp) , \nonumber\\
\hat S_2(\vp) &= -i\big(a^\dagger_1(\vp)a_2(\vp) - a^\dagger_2(\vp)a_1(\vp)\big) , \nonumber\\
\hat S_3(\vp) &= a^\dagger_1(\vp)a_1(\vp) - a^\dagger_2(\vp)a_2(\vp) .
\end{align}
The operator $\hat S_3(\vp)$ is the basic polarization observable
\eqref{eq:PM9}, and $\hat S_1(\vp)$ arises from the rotation of the
polarization filter. The operator $\hat S_2$ measures circular polarization.
Its eigenstates in the one-photon sector are created by
$(a_1^\dagger\pm ia_2^\dagger)/\sqrt2$. The polarization operator
$\hat P_\beta(-\vp)$ for the second photon obtains by replacing $\vp\to-\vp$.

The Stokes operators are bilinears of the annihilation and creation operators
of the same mode. The commutation relations \eqref{eq:aacommdiscrete} imply
the $SU(2)$-algebra (Schwinger representation)
\begin{equation}\label{eq:SU2}
\big[\hat S_i(\vp),\hat S_j(\vp\,')\big]
= 2i\,\delta_{\vp\vp\,'}\,\varepsilon_{ijk}\,\hat S_k(\vp) .
\end{equation}
This holds on the whole Fock space. On the one-photon sector of a given mode
the Stokes operators are represented by the Pauli matrices,
$\hat S_i\to\tau_i$, and $\hat P_\alpha(\vp)$ becomes the real symmetric matrix
\eqref{eq:PQ4},
\begin{equation}\label{eq:Ralpha}
R(\alpha) = \cos2\alpha\,\tau_3 + \sin2\alpha\,\tau_1
= \begin{pmatrix}\cos2\alpha & \sin2\alpha\\ \sin2\alpha & -\cos2\alpha\end{pmatrix} ,
\end{equation}
with $R^2(\alpha)=1$ and eigenvalues $\pm1$.

In terms of the Stokes operators the non-commutativity \eqref{eq:PM15},
\eqref{eq:PM16} of polarization operators for the same photon takes the
compact form
\begin{equation}\label{eq:PPcomm}
\big[\hat P_\alpha(\vp),\hat P_{\gamma}(\vp)\big]
= 2i\,\sin2(\gamma-\alpha)\,\hat S_2(\vp) .
\end{equation}
Two polarization observables for the same photon commute only if the
polarizer axes are parallel or orthogonal. In contrast, polarization
operators for the two different photons always commute,
\begin{equation}\label{eq:PABcomm}
\big[\hat P_\alpha(\vp),\hat P_\beta(-\vp)\big] = 0 .
\end{equation}
Furthermore, all Stokes operators commute with the quantum Hamiltonian
$H^{(q)}$ in eq.~\eqref{eq:Hqn}, since $H^{(q)}$ depends only on the total
occupation number $\hat n_1(\vp)+\hat n_2(\vp)$ of each mode. The polarization
of free photons is conserved, and the correlations computed below do not
depend on time.

It is instructive to express the Stokes operators in terms of the field
operators $\hat\varphi_k$ and $\hat\eta_k$. With eq.~\eqref{eq:aadagger} one
finds
\begin{align}\label{eq:Stokesfields}
\hat S_1(\vp) &= p\,\hat\varphi_1(\vp)\hat\varphi_2(\vp)
+ \frac1p\,\hat\eta_1(\vp)\hat\eta_2(\vp) , \nonumber\\
\hat S_2(\vp) &= \hat\varphi_1(\vp)\hat\eta_2(\vp)
- \hat\varphi_2(\vp)\hat\eta_1(\vp) , \nonumber\\
\hat S_3(\vp) &= \frac p2\big(\hat\varphi_1^2(\vp)-\hat\varphi_2^2(\vp)\big)
+ \frac1{2p}\big(\hat\eta_1^2(\vp)-\hat\eta_2^2(\vp)\big) .
\end{align}
The constant contributions from the commutator $[\hat\varphi_k,\hat\eta_k]$
cancel in $\hat S_3$. The circular polarization $\hat S_2$ is the generator of
internal rotations in the plane transverse to $\vp$, i.e.\ the helicity. It
is the contribution of the mode $\vp$ to the generator $\hat L^{(q)}_3$ in
eq.~\eqref{eq:R5}. The
non-commutativity \eqref{eq:PPcomm} has its origin in the canonical
commutation relation between $\hat\varphi_k$ and $\hat\eta_k$. In the original
basis \eqref{eq:operators} of the classical wave function the field operators
are $\hat\varphi_k=\hat\sigma_k+\frac12\hat\zeta_k$ and
$\hat\eta_k=\hat\pi_k+\frac12\hat\gamma_k$. The polarization operators are
therefore second order differential operators acting on $q(\sigma,\pi)$. They
are not multiplication operators by functions of the classical fields
$\sigma_k(\vp)$ and $\pi_k(\vp)$.

The explicit form of the Stokes operators in the $(\sigma,\pi)$ basis reads
(for $\vp=(0,0,p_3)$)
\begin{align}\label{eq:Stokessigmapi}
\hat S_1(\vp) ={}& p\,\sigma_1\sigma_2 + \frac1p\,\pi_1\pi_2
- \frac p4\,\frac{\partial^2}{\partial\pi_1\partial\pi_2}
- \frac1{4p}\,\frac{\partial^2}{\partial\sigma_1\partial\sigma_2}
\nonumber\\
&+ \frac i2\,p\Big(\sigma_1\frac{\partial}{\partial\pi_2}
+ \sigma_2\frac{\partial}{\partial\pi_1}\Big)
\nonumber\\
&- \frac i{2p}\Big(\pi_1\frac{\partial}{\partial\sigma_2}
+ \pi_2\frac{\partial}{\partial\sigma_1}\Big) ,
\nonumber\\
\hat S_3(\vp) ={}& \frac p2\big(\sigma_1^2-\sigma_2^2\big)
+ \frac1{2p}\big(\pi_1^2-\pi_2^2\big)
\nonumber\\
&- \frac p8\Big(\frac{\partial^2}{\partial\pi_1^2}-\frac{\partial^2}{\partial\pi_2^2}\Big)
- \frac1{8p}\Big(\frac{\partial^2}{\partial\sigma_1^2}
-\frac{\partial^2}{\partial\sigma_2^2}\Big)
\nonumber\\
&+ \frac i2\,p\Big(\sigma_1\frac{\partial}{\partial\pi_1}
- \sigma_2\frac{\partial}{\partial\pi_2}\Big)
\nonumber\\
&- \frac i{2p}\Big(\pi_1\frac{\partial}{\partial\sigma_1}
- \pi_2\frac{\partial}{\partial\sigma_2}\Big) ,
\end{align}
where $\sigma_k\equiv\sigma_k(\vp)$, $\pi_k\equiv\pi_k(\vp)$. For the
correlated two-photon state \eqref{eq:psi2ex}, \eqref{eq:q2} we can therefore
evaluate the expectation value of a product $P(\alpha)P(\beta)$ by
\begin{equation}\label{eq:PPq2}
\langle P(\alpha)P(\beta)\rangle = \int\mathcal{D}\sigma\,\mathcal{D}\pi\;
q_2(\sigma,\pi)\,\hat P(\alpha)\hat P(\beta)\,q_2(\sigma,\pi) .
\end{equation}
This demonstrates how expectation values for products of two polarization
observables can be directly computed from the classical wave function $q_2$
for the correlated two-photon state. Since $\hat P(\alpha)$ and $\hat P(\beta)$ act on
different momentum modes they commute. There is no ordering problem for the
derivatives.

\subsection{Polarization correlations}

The expectation value \eqref{eq:PPq2} defines a correlation function for the
polarizations of two photons with opposite momentum. The two photons can be
measured by separate polarization detectors. The measurement of the photon
$A$ does not influence the measurement of the photon $B$. For arbitrary
angles $\alpha$ and $\beta$ of the polarization filters one can measure the
probabilities
\begin{equation}\label{eq:wab}
w_{\alpha\beta} = \langle n_\alpha n_\beta\rangle
= \frac14\big\langle\big(1+P(\alpha)\big)\big(1+P(\beta)\big)\big\rangle ,
\end{equation}
that one detects a photon in each detector, and similarly
$w_\alpha=\langle n_\alpha\rangle$ and $w_\beta=\langle n_\beta\rangle$ that
one detects a photon of type $A$ or $B$. The expectation values as
$\langle n_\alpha n_\beta\rangle$ define correlation functions that can be
measured. Since the polarization operators for the two different photons
commute there is no ambiguity in the definition of these correlation
functions.

Nothing essential changes if one uses wave packets instead of sharp momenta.
In this case one replaces $a_k(\vp)$ by
$a_k(f_A)=\int_{p'}f_A^*(\vp\,')\,a_k(\vp\,')$ for photon $A$, with a
normalized wave packet $f_A$ centered around $\vp$, and similarly
$a_k(f_B)$ for photon $B$, with $f_B$ centered around $-\vp$. For
non-overlapping wave packets the commutation relations
$[a_k(f_A),a^\dagger_l(f_A)]=\delta_{kl}$,
$[a_k(f_A),a^\dagger_l(f_B)]=0$ take the same form as
eq.~\eqref{eq:aacommdiscrete}. The Stokes operators and polarization
operators are constructed from $a_k(f_A)$ and $a_k(f_B)$ exactly as in
eqs.~\eqref{eq:PalphaStokes}, \eqref{eq:Stokes}. They obey the same algebra
\eqref{eq:SU2}, and on the one-photon sectors they are again represented by
the matrices $R(\alpha)$ and $R(\beta)$. (For wave packets narrow in momentum
space the transverse polarization directions $1$ and $2$ can be taken fixed.)
The correlation functions are therefore the same as for sharp momenta. Wave
packets can be localized in space, such that the two polarization
measurements can be performed by detectors at different locations, without
the factors $\delta(0)$ arising for sharp momenta.

Generalizing eqs.~\eqref{eq:PQ2}, \eqref{eq:PQ3} to two photons, the
polarization operator $\hat P_\alpha(\vp)$ acts only on the first index of
$c_{kl}$, and $\hat P_\beta(-\vp)$ only on the second one,
\begin{equation}\label{eq:PonC}
\hat P_\alpha(\vp)\,\hat P_\beta(-\vp)\,\psi_2
= \sum_{k,l}\big(R(\alpha)\,C\,R(\beta)\big)_{kl}\,|kl\rangle .
\end{equation}
The expectation value of the product of the two polarization observables,
\begin{equation}\label{eq:Edef}
E(\alpha,\beta) = \big\langle\psi_2\big|\hat P_\alpha(\vp)\hat P_\beta(-\vp)\big|\psi_2\big\rangle ,
\end{equation}
is therefore given by the simple formula
\begin{equation}\label{eq:EC}
E(\alpha,\beta) = \mathrm{tr}\big(C^\dagger R(\alpha)\,C\,R(\beta)\big) .
\end{equation}
Similarly, the polarization of photon $A$ alone has the expectation value
\begin{equation}\label{eq:PA}
\langle\hat P_\alpha(\vp)\rangle = \mathrm{tr}\big(C^\dagger R(\alpha)\,C\big)
= \mathrm{tr}\big(\rho_A R(\alpha)\big) , \qquad \rho_A = CC^\dagger ,
\end{equation}
and correspondingly for photon $B$ with $\rho_B = (C^\dagger C)^T$.

For the maximally entangled states \eqref{eq:examples} one obtains
\begin{equation}\label{eq:Emax}
E_\pm(\alpha,\beta) = \pm\cos2(\alpha-\beta) , \qquad
\langle\hat P_\alpha(\vp)\rangle = \langle\hat P_\beta(-\vp)\rangle = 0 .
\end{equation}
Each photon separately is completely unpolarized: for any orientation of a
single polarizer the photon is transmitted with probability $1/2$. The
polarizations of the two photons are nevertheless perfectly correlated for
$C_+$ (perfectly anticorrelated for $C_-$) whenever the two polarizers are
parallel, $\alpha=\beta$, independently of the common angle $\alpha$. In
terms of probabilities, the probability that both photons are transmitted is
$\frac14\big(1+E(\alpha,\beta)\big)=\frac12\cos^2(\alpha-\beta)$ for $C_+$.
For a product state, $c_{kl}=a_kb_l$, the correlation factorizes,
$E(\alpha,\beta)=\langle\hat P_\alpha(\vp)\rangle\langle\hat P_\beta(-\vp)\rangle$.

One verifies that eq.~\eqref{eq:Emax} obtains from eq.~\eqref{eq:PPq2} by
using the explicit form \eqref{eq:Stokessigmapi} of the Stokes operators.
In the $(\sigma,\pi)$ basis each Stokes operator decomposes as
$\hat S = R + iX$, with $R$ a real symmetric operator (multiplication and
second derivatives) and $X$ a real antisymmetric first-order differential
operator. For real $q$ one has $\int q\,X\,q = 0$, such that expectation
values are real. For the product of two commuting operators the real part
reads $R_AR_B - X_AX_B$, and the first-order terms contribute to the
correlation function. For the state \eqref{eq:psi2ex} they account for half
of the correlation $\cos2(\alpha-\beta)$. Without them $|S|\le\sqrt2$, and
Bell's inequalities would not be violated.

\subsection{CHSH inequality}

Consider two orientations $\alpha,\alpha'$ for the polarizer of photon $A$
and two orientations $\beta,\beta'$ for photon $B$. The CHSH combination
of correlations is
\begin{equation}\label{eq:Sdef}
S = E(\alpha,\beta) + E(\alpha,\beta') + E(\alpha',\beta) - E(\alpha',\beta') .
\end{equation}
If the outcomes of the polarization measurements were given by
predetermined values $A_\alpha(\lambda),A_{\alpha'}(\lambda)$ and
$B_\beta(\lambda),B_{\beta'}(\lambda)$, all equal to $\pm1$, with $\lambda$
distributed according to some probability distribution, and if the outcome
for photon $A$ did not depend on the setting for photon $B$ and vice versa,
one would have
\begin{equation}\label{eq:CHSH}
|S| \le 2 .
\end{equation}
This follows from
$A_\alpha(B_\beta+B_{\beta'})+A_{\alpha'}(B_\beta-B_{\beta'})=\pm2$ for every
$\lambda$.

For $C_+$ we choose
\begin{equation}\label{eq:angles}
\alpha=0 ,\quad \alpha'=\frac\pi4 ,\quad \beta=\frac\pi8 ,\quad \beta'=-\frac\pi8 ,
\end{equation}
for which $\sin2(\alpha'-\alpha)\sin2(\beta'-\beta)=-1$. Explicitly, eq.~\eqref{eq:Sdef} gives
\begin{equation}\label{eq:Smax}
S = 3\cos\frac\pi4 - \cos\frac{3\pi}4 = 2\sqrt2 .
\end{equation}
For $C_-$ one has $E_-=-E_+$, such that the same settings yield
$S=-2\sqrt2$. In both cases the CHSH inequality \eqref{eq:CHSH} is violated by the factor
$\sqrt2$.

More generally, in the quantum field theory for photons $S$ is the expectation value of the
Bell operator
\begin{align}\label{eq:Bellop}
\hat{\mathcal B} = {}&\hat P_\alpha(\vp)\big(\hat P_\beta(-\vp)+\hat P_{\beta'}(-\vp)\big)
\nonumber\\
&+ \hat P_{\alpha'}(\vp)\big(\hat P_\beta(-\vp)-\hat P_{\beta'}(-\vp)\big) .
\end{align}
Using $\hat P^2=1$ on the two-photon subspace and eq.~\eqref{eq:PABcomm}, its
square is
\begin{equation}\label{eq:Bellsquare}
\hat{\mathcal B}^2 = 4 - \big[\hat P_\alpha(\vp),\hat P_{\alpha'}(\vp)\big]
\big[\hat P_\beta(-\vp),\hat P_{\beta'}(-\vp)\big] .
\end{equation}
This identity shows in the clearest way the role of non-commutativity. If
the two polarization observables for one of the photons commuted, one would
have $\hat{\mathcal B}^2=4$, and therefore $|S|\le2$ for every state. With
eq.~\eqref{eq:PPcomm} one finds
\begin{equation}\label{eq:Bellsquare2}
\hat{\mathcal B}^2 = 4 + 4\sin2(\alpha'-\alpha)\,\sin2(\beta'-\beta)\,
\hat S_2(\vp)\,\hat S_2(-\vp) .
\end{equation}
The eigenvalues of $\hat S_2(\vp)\hat S_2(-\vp)$ are $\pm1$, such that
$\hat{\mathcal B}^2\le8$ and
\begin{equation}\label{eq:Tsirelson}
|S| \le 2\sqrt2 .
\end{equation}
This is Tsirelson's bound~\cite{TSI}.

The maximally entangled states \eqref{eq:examples} saturate this bound. Both
$C_+$ and $C_-$ are eigenstates of $\hat S_2(\vp)\hat S_2(-\vp)$ with
eigenvalue $-1$, since
$\tau_2 C_\pm\tau_2^T = -C_\pm$. For the angles \eqref{eq:angles} this
yields $\hat{\mathcal B}^2\psi_2=8\,\psi_2$.

For a general two-photon state the maximal value of $|S|$, optimized over
all settings, follows from the correlation matrix
$T_{ij}=\langle\psi_2|\hat S_i(\vp)\hat S_j(-\vp)|\psi_2\rangle
=\mathrm{tr}(C^\dagger\tau_iC\tau_j^T)$~\cite{HOR}. If arbitrary analyzers on the
Poincar\'e sphere are admitted, with observables $\vec n\cdot\hat{\vec S}$
realized by combinations of wave plates and linear polarizers, one finds for
pure states~\cite{GIS,HOR}
\begin{equation}\label{eq:Sgeneral}
S_{\max} = 2\sqrt{1+\mathcal{C}^2} .
\end{equation}
Every entangled two-photon state, $\det C\neq0$, violates the CHSH
inequality, while product states do not. If $C$ is real up to an overall
phase, linear polarizers are sufficient. In this case $C$ can be brought to
the diagonal form $\mathrm{diag}(\cos\theta,\sin\theta)$ by choosing
appropriate linear polarization bases for the two photons. With respect to
these bases the relevant correlations are $T_{33}=1$ and
$T_{11}=\pm\sin2\theta=\pm\mathcal C$, and eq.~\eqref{eq:Sgeneral} is reached with
linear polarizers only.

\subsection{Classical probability distribution and Bell's theorem}

By the construction of sect.~\ref{sec:prob} each of the two-photon states
\eqref{eq:psi2C} defines a real classical wave function $q(\sigma,\pi)$ and
therefore a positive normalized probability distribution
$w(\sigma,\pi)=q^2(\sigma,\pi)$ for the classical fields $A^T_k$ and
$\partial_tA^T_k$. The time evolution of $w$ is given by the Liouville
equation for the classical Maxwell equations. Nevertheless, the polarization
correlations \eqref{eq:EC} violate Bell's inequalities. This seems to be in
conflict with Bell's theorem. The resolution lies in the observables.

As discussed in sect.~\ref{sec:polmeas}, expectation values are computed as
bilinears in the classical wave function. For the pure states
\eqref{eq:purestate} this can be made explicit. Since
the functional Fourier transform \eqref{eq:FFT} is unitary for a suitable
normalization of the functional measure, and since
$\langle\tilde\psi|\hat A|\tilde\psi\rangle
=\langle\psi|\hat A|\psi\rangle\langle\psi|\psi\rangle$ for an operator
$\hat A$ acting on $\varphi$, one has
\begin{equation}\label{eq:Aq}
\langle\hat A\rangle = \int\mathcal{D}\sigma\,\mathcal{D}\pi\;
q(\sigma,\pi)\,\hat A\,q(\sigma,\pi) ,
\end{equation}
where $\hat A$ is expressed in terms of $\hat\sigma,\hat\pi,\hat\gamma,\hat\zeta$.
For the polarization operators \eqref{eq:Stokesfields} the right-hand side of
eq.~\eqref{eq:Aq} involves derivatives of $q$ with respect to $\sigma$ and
$\pi$. These expectation values are not of the form
$\int w(\sigma,\pi)A(\sigma,\pi)$ for some function $A(\sigma,\pi)$ on the
phase space of classical fields. The probabilistic information needed for
their computation is contained in $q$.

Bell's inequalities hold for correlations of the type
$\langle A_\alpha B_\beta\rangle_{\rm cl}=\int w\,A_\alpha B_\beta$, where
$A_\alpha(\sigma,\pi)$ and $B_\beta(\sigma,\pi)$ are functions on phase space
taking values $\pm1$. For such classical correlations all four observables
$A_\alpha,A_{\alpha'},B_\beta,B_{\beta'}$ have simultaneous values in every
state of the ensemble, and $|S|\le2$ follows. The polarization observables
of the quantum subsystem are represented by non-commuting operators
\eqref{eq:PPcomm}. There is no joint probability distribution for the values
of $\hat P_\alpha(\vp)$ and $\hat P_{\alpha'}(\vp)$, and the derivation of
eq.~\eqref{eq:CHSH} does not apply. The existence of a positive classical
probability distribution $w(\sigma,\pi)$ for the fields is therefore not in
conflict with the violation of Bell's inequalities by the photon
polarization correlations.

The decisive question is then which correlation describes the outcomes of
real measurements with polarizers. In sect.~\ref{sec:polmeas} we have argued
that an ideal polarization detector is represented by the operator $\hat n_1$
acting on the classical wave function, with Born's rule for the probability
that it fires, and that the polarization observables $P(\alpha)$ for
arbitrary angles follow from rotations. With this measurement rule the
correlations of two polarization measurements on the two photons are the
correlations \eqref{eq:EC}, not classical correlations of the type
$\int w\,A_\alpha B_\beta$. The polarization observables are statistical observables
in the sense of sect.~\ref{sec:polmeas}, since their operators involve
derivatives with respect to $\sigma$ and $\pi$.

\section{Statistical observables}\label{sec:statobs}

We have seen how entangled two-photon states can be described by a classical
probability distribution. In this section we discuss the question of possible
measurements more generally. Every theory needs some prescription for the
computation of expectation values, and some assumptions how these expectation
values are related to real ideal measurements.

We first have to ask what can be measured in our type of classical field
theory. Assume that we formulate our model on a lattice, with lattice
distance smaller than the Planck length. It is obvious that one cannot
measure precisely the field variables $\sigma(\vx)$. This would require some
detector with resolution smaller than the Planck length. What can be measured
are more macroscopic observables. We should understand the fields
$\sigma(\vx)$ as the microscopic variables which are used to formulate the
theory. They should not be associated directly with the observables for real
measurements.

The macroscopic observables typically involve properties of the probability
distribution. An example are expectation values $\bar A(\vx)$ in local regions
centered around $\vx$. These are not simply functions of $\sigma(\vx)$. They
involve properties of the local probability distribution. We can then define
expectation values over larger regions $\langle\bar A(\vx)\rangle$, or
correlations $G(\vx,\vec y)=\langle\bar A(\vx)\bar A(\vec y)\rangle$. Such
correlations are typically no longer classical observables in the sense that
a function $B[\tilde\sigma]$ exists such that
$G(\vx,\vec y)=\int\mathcal D\tilde\sigma\,w(\tilde\sigma)B(\tilde\sigma)$.
Local properties of the probability distribution around $\vx$ and $\vec y$
enter in a more complex way. For a description of real measurements we
should therefore consider a class of possible observables which include
statistical observables measuring at least partly properties of the
probabilistic information. One is actually doing this in practice for typical
many body systems if one measures temperature correlations or correlations
for the local magnetization. The conceptual issue may be somewhat obscured, however, if
one still could have access, in principle, to the microscopic variables as
positions and momenta of atoms. For a field theory with resolution below the
Planck length the necessity of statistical observables is rather obvious.

The properties of the probabilistic information may involve its roughness, or
the question how strongly it changes if the microscopic field is varied on
microscopic scales. Statistical observables depending on the roughness
typically involve derivatives with respect to the fields. The formulation of
the time-local probabilistic information in terms of the classical wave
function offers the possibility of a simple description of a large class of
statistical observables. For the class of ``linear observables'' the
expectation value of $A$ can be computed by associating to $A$ an operator
$\hat A$ and forming an expression bilinear in $q$
\begin{equation}\label{eq:S01}
\langle A\rangle = \int\mathcal D\tilde\sigma\,q(\tilde\sigma)\,\hat A\,
q(\tilde\sigma) .
\end{equation}
Here $\hat A$ can involve derivatives with respect to the fields
$\partial/\partial\tilde\sigma(\vx)$. No similar expressions linear in $w$ can
be formulated, since
$\int\mathcal D\tilde\sigma\,\frac{\partial}{\partial\tilde\sigma(\vx)}\,
w(\tilde\sigma)$ is a total derivative. The linear observables \eqref{eq:S01}
are not the most general statistical observables. For example, temperature or
pressure do not admit a simple expression \eqref{eq:S01}. The linear
observables will be, however, sufficient for the purpose of this paper.

The most important statistical observables for the present paper are the
photon numbers $n_k(\vp)$. They involve derivatives with respect to $\sigma$
and $\pi$ and are therefore statistical observables. The very notion of
subsystems with a fixed photon number is based on the observables
$n_k(\vp)$. If a measurement apparatus can detect single photons, it is
naturally related to the observables $n_k(\vp)$. This is covered by our basic
assumption on the polarization detector for a single photon being based on
$n_1(\vp)$. The conclusion of this short section is simple. There is no
justification to restrict the possible observables to classical observables
which are functions of the classical fields. A large class of linear
observables includes derivative observables which play an important role as
conserved quantities.

\section{Conclusions}\label{sec:concl}

We have presented a classical probability distribution for electromagnetic
fields which represents states with two correlated photons
\eqref{eq:q2}. Polarization measurements for the two photons are described
by correlation functions which violate Bell's inequalities. The
experimentally observed violation of Bell's inequalities is not a sign that
nature cannot be described in terms of classical probabilities. It rather
points to the presence of important statistical observables for classical
probabilistic systems. They measure partially properties of the probabilistic
information.

Our argument is based on two simple assumptions. The first states that
electromagnetism in vacuum is described by a classical field equation for the
gauge field which obeys the usual Lorentz- and gauge invariance. The
description is probabilistic in the sense that probabilistic initial
conditions for field configurations are specified. Probabilities are
considered here as the fundamental description of nature, not as an effect of
missing knowledge about an otherwise deterministic world~\cite{CWPW}. Second,
we assume that an ideal polarization detector for a single photon measures
the probability of finding a photon according to the mean value of the photon
number computed from the classical probability distribution. The photon
number is a conserved statistical observable. Its expectation value is
computed from an associated operator acting linearly on the classical wave
function, which is the (real) root of the probability distribution. From
these two assumptions we have derived the violation of Bell's inequalities
for correlations in a two-photon system.

Our two assumptions reflect the basics of any measurement process. A
theoretical framework has to compute expectation values of observables, and one needs an
assumption how a given measurement apparatus is represented by an observable.
Our approach assumes implicitly that the photons observed in nature
correspond to the multi-photon states present in the probabilistic classical
setting. The classical statistical setting indeed exhibits conserved photon
numbers, such that probability distributions or classical wave functions with
a fixed photon number form closed subsystems. The conserved photon number is,
however, a statistical observable which cannot be expressed as a function of
the classical fields. It describes properties of the probabilistic
information -- namely the possibility of closed multi-photon subsystems. The
conserved photon number is crucial for our description of polarization
measurements. This underlines the key role of statistical observables. The
photon numbers cannot be expressed as functions of the classical fields.

More generally, statistical observables are not simply functions of the
classical fields. The operators associated to the statistical observables
involve derivatives with respect to the fields. As a consequence,
simultaneous probabilities for arbitrary pairs of observables do not exist --
statistical observables are characterized by ``incomplete
statistics''~\cite{CWIS1,CWIS2,CWPW}. For two given angles $\alpha$ and
$\beta$ of the polarization filters for the two photons simultaneous
probabilities for the given pairs of polarization observables do exist.
What does not exist are simultaneous probabilities for further polarization
directions $\alpha'$ and $\beta'$. These would be required for a proof of the
CHSH-inequality. Since this essential assumption for the proof is not met,
suitable combinations of polarization correlations can and do violate the
CHSH-bound.

The absence of simultaneous probabilities in incomplete statistics is
directly related to the non-commuting properties of operators. We have
explicitly constructed the quantum field operators for the electromagnetic
field. They obey non-trivial commutation relations. The subsystem of
observables based on these quantum fields forms a quantum field theory for
free photons. This presence of a quantum field theory as a subsystem of the
probabilistic classical field theory is the deeper reason why a quantum
description of correlated photons can be embedded in a classical
probabilistic setting.

We have only described a classical field theory for free photons. This is
sufficient for an explanation of the violation of Bell's inequalities for
correlations of two photons. Real quantum electrodynamics needs the
interaction of photons with charged particles. This is required both for the
preparation of the entangled two-photon state, and for the real measurement
process. In this respect our treatment is not complete. It is sufficient,
however, to formulate the conceptual basis for ``quantum correlations in
classical statistics''. And it demonstrates explicitly that the violation of
Bell's inequalities is not incompatible with a classical probabilistic
description.

Can probabilistic classical fields describe the photons observed in nature?
This is to a large extent a dynamical question. In our setting a key feature
is the quantum vacuum. Photons are excitations of the vacuum with positive
quantum energy. The question arises if this quantum vacuum is a result of the
dynamics in a system with suitable interactions. The quantum vacuum has a
non-zero expectation value of the classical field energy~\cite{CWQV}. A
competing state is the classical thermal equilibrium with the same mean
classical energy. The classical thermal equilibrium shows the Rayleigh--Jeans
divergence, while the quantum vacuum has only a divergent ground state energy
which drops out in the probability distribution.

The equal-time correlation functions for classical fields differ for the
quantum vacuum and the classical thermal equilibrium with the same mean
classical field energy. There exists already a substantial body of work
investigating the time evolution of field
correlations~\cite{PSTA,BMS95,DTS,YK,AS1,CWET1,CWET2,NAC,BUJA,AS2,BOWE,ABW,CKH,AB2002,MT04,BBW04,FGL,GLV,BBG,EGW,BER,TSE,OZB}.
It is an interesting question if these methods can find a suitable model with
interactions for which the correlation functions approach for asymptotic
times the ones for the quantum vacuum. Suitable charged fields may act
effectively as stochastic forces on the electromagnetic field and there may
be a connection to stochastic
electrodynamics~\cite{MAR1,BOY1,BOY2,BOY3,QDICE}. If one can find a model
with interactions for which the quantum ground state (or an approximation to
it) is realized naturally, the present work may not only provide a
counter-example for no-go theorems based on Bell's inequality. It could
describe the observed photons in nature.

\begin{acknowledgments}
The author has worked with the AI Claude for literature search, formulation
of sect.~\ref{sec:Bell}, derivation and check of some formulae, figure, and typesetting.
\end{acknowledgments}

\nocite{*}
\bibliography{refs}

\end{document}